\documentclass{article}
\usepackage{amsmath,amssymb}
\usepackage{graphicx}
\usepackage[normalem]{ulem}
\usepackage{xcolor}
\usepackage{empheq}
\usepackage[numbers]{natbib}
\usepackage{graphicx}

\usepackage{url} 
\usepackage{lineno}

\def\be{\begin{equation}}
\def\ee{\end{equation}}

\title{Evaluating the incompleteness magnitude using an unbiased estimate of the $b$ value.}

\author{\large{C. Godano$^{1,2}$, G. Petrillo $^3$, E. Lippiello$^{1}$}}
  \date{\small{$^{1}$Department of Mathematics and Physics,
 Universit\'a della Campania "Luigi Vanvitelli", Caserta, Italy.}\\
 \small{$^{2}$INGV-Osservatorio Vesuviano, Napoli, Italy} \\
 \small{$^3$ The Institute of Statistical Mathematics, Research Organization of Information and Systems, Tokyo, Japan}
}

\begin{document}

\label{firstpage}

\maketitle

\section*{Abstract}
  The evaluation of the $b$ value of the Gutenberg-Richter (GR) law, for a sample composed of $n$ earthquakes, presents a systematic positive bias $\delta b$ which is proportional to $1/n$, as already observed by Ogata \& Yamashina (1986). In this study we show how to incorporate in $\delta b$ the bias introduced by deviations from the GR law. More precisely we show that $\delta b$ is proportional to the square of the variability coefficient $CV$, defined as the ratio between {the standard deviation of the magnitude distribution and its mean value.} When the magnitude distribution follows the GR law $CV=1$ and this allows us to introduce a new procedure, based on the dependence of $b$ on $n$, which allows us to {identify} the incompleteness magnitude $m_c$ as the threshold magnitude leading to $CV=1$. The method is tested on synthetic catalogs and it is applied to estimate $m_c$ in Southern California, Japan and New Zealand.

\section{Introduction}

The probability to observe an earthquake of magnitude $m$ is well described by the Gutenberg and Richter (GR) \cite{GR44} law 
\begin{equation}
p(m)=b\ln(10)10^{-b(m-m_c)}
\label{gr}
\end{equation}

being $b$ the scaling parameter and $m_c$ a reference magnitude.

The evaluation of $b$ has received an increasing interest over years mostly because of experimental results  which indicate that it is inversely correlated to stress 
state \cite{Sch68, Wys73, Ami03, Gw10,Sch15} and directly correlated to the material heterogeneity \cite{Mog62}. Recently a unifying empirical framework is proposed to link the $b$ value with Anderson’s faulting theory and differential stress \cite{PSTRWGV19}.
These results are also in agreement with recent numerical \cite{LPLR19,PLLR20,LPLR21} and experimental findings \cite{GTPBL22}, indicating that the $b$ value decreases due to a stress concentration induced by afterslip dynamics. Within the hypothesis that $b$ is related to the stress state, one can use its evolution in time and space to identify stress changes induced by big foreshocks and in this way to identify eventual precursory patterns. This idea 
has therefore promoted a lot of effort in the identification of spatio-temporal variations of the $b$-value \cite{WW97,WW02,Gw10,NHOK12,TWM14,TEWW15,GW19,Nan20,TGRGCODC21,GDBPP21}.
Nevertheless, much attention must be devoted to properly discriminate between genuine and spurious $b$-value variations \cite{MSST19}. 
A first problem is caused by the limited number $n$  of earthquakes used to measure the $b$-value. In particular, \cite{OY86} have shown that the finiteness of the data set introduces a systematic positive bias in the evaluation of $b$, which is proportional to $1/n$. Most studies ignore this systematic effect since it is of the same order of the $b$-value variance, which is also of order $1/n$ \cite{SB82}.
A second problem is related to the difficulty of identifying earthquakes
when the signal to noise ratio is too small \cite{Amo07}. This makes seismic catalogs incomplete, in the sense that many small earthquakes are not reported. The degree of incompleteness depends on the capability to filter noise and on the distance between the earthquake epicenter and the seismic stations necessary to trigger an event declaration in a catalog \cite{SW08,Mig12}. At the same time, incompleteness becomes particularly relevant in the first part of aftershock sequences because of  coda waves of large earthquakes, and their aftershocks, that cover
subsequent smaller earthquakes \cite{Kag04,HKJ06,PVIH07,LdAG08,LGdA12,OOHA13,dAGGL16,OOSESA16,Hai16,Hai16a,ZOW17,dAGL18}. This problem is usually solved by restricting the analysis to earthquakes with magnitude larger than a completeness magnitude $m_c$, defined as the magnitude above which all the earthquakes 
occurred in a given area are reasonably recorded and reported in the catalog \cite{RS01}.

A precise estimate of $m_c$ is crucial for a correct evaluation of the $b$ value. Indeed an underestimate of $m_c$ will reflect into an underestimate of the $b$ value. Conversely, an overestimate of $m_c$ will imply a loss of information 
reducing the magnitude interval over which $b$ is estimated. Several methods have been proposed for estimating the $m_c$ 
value and here we quote, as representative, the maximum curvature technique \cite{WW00} based on the maximum value of 
the non cumulative GR distribution; the test of fit goodness \cite{WW00} that performs an exponential fit of the GR as a 
function of a threshold magnitude $m_{th}$ and selects $m_c$ as the first $m_{th}$ for which an accurate fit is obtained; the 
$b$-value stability approach \cite{CG02} which chooses $m_c$ as the $m_{th}$ value at which the estimated $b$ value becomes 
stable; the entire magnitude range method \cite{OK93} using the entire range of magnitude when the GR distribution is 
multiplied by a complementary error function; the harmonic mean based method introduced by \cite{God17} and also \cite{GP22}.

In this study we show that deviations on the $b$ estimate caused by a finite $n$ are related to deviations induced by  an incorrect estimate of $m_c$. In particular, by the use of the central limit theorem, we obtain an expression of the corrections to the $b$ value caused by the finiteness of the sample and by deviations from the GR law. This expression is used as a novel tool to identify $m_c$.

\section{An unbiased estimate of the \lowercase{$b$} value}

Assuming that magnitude distribution obeys the GR law Eq.(\ref{gr}) for magnitudes larger than a reference value $m_{th}$, in the limit $n\to \infty$, from likelihood maximization one obtains \cite{Aki65}
\be
b_{\infty}=\frac{1}{\ln(10)(\langle m\rangle-m_{th})},
\label{ml}
\ee
where $\langle m\rangle$ is the average magnitude in the data set.

In a data set $\{m_1,...,m_n\}$ containing a finite number $n$ of events, Eq.(\ref{ml}) is replaced by 
\begin{equation}
  b_n=\frac{1}{\ln(10) X_n},
  \label{ml2}
\end{equation}
 where
 \begin{equation}
X_n=\frac{1}{n}\Sigma_{i=1}^n \left(m_i-m_{th}\right).
\end{equation}

 The difference between the estimate of $b$, obtained from Eq.(\ref{ml2}), from the exact $b$ value $b_{\infty}$ has been rigorously obtained by \cite{OY86}, who demonstrate that
  the average value of $b$ in a finite set is given by
\begin{equation}
 \langle b_n \rangle=b_{\infty}\frac{n}{n-1}= b_{\infty}\left(1+\frac{1}{n}\right)+ {\cal O}\left(1/n^2\right).
  \label{oy}
\end{equation}
Eq.(\ref{oy}) was obtained using the previous result by \cite{Uts66} for the whole functional form of the distribution of the $b$ value, for a finite data set distributed according to the GR law.

\subsection{Analytical derivation of Eq.(\ref{oy}) from the central limit theorem}

{ We present the analytical derivation of Eq.(\ref{oy}). We begin by considering the definition of $b_{\infty}$ from Eq.(\ref{ml2}), assuming that magnitudes are i.i.d. variables without specifying their distribution. Only in the final step do we specify our results for the case of magnitudes distributed according to the GR law, with $b_{\infty}$ coinciding with the $b$-value. This approach allows us to obtain a general expression for $b_{\infty}$ that accounts for deviations from the GR law.}

  According to the central limit theorem, for sufficiently large $n$, the quantity  $ z_n=\sqrt n \left(X_n-\mu\right)$ follows a 
normal distribution with mean $0$ and standard deviation $\sigma/\sqrt{n}$,
where $\mu=\langle m_i-m_{th}\rangle$ is the average value of $\left(m_i-m_{th}\right)$ and 
$\sigma=\left \langle \left(m_i-m_{th}-\mu\right)^2 \right \rangle$ represents its standard deviation.
Accordingly, the average value of $b_n$, defined in Eq.(\ref{ml2}), is given  by

\begin{eqnarray}
 \langle b_n \rangle &= &\frac{1}{\ln(10)\sqrt{2\pi \sigma^2} }\int_{-\infty}^{\infty}dz_n\frac{1}{\mu+z_n/\sqrt {n}} \exp^{-\frac{z_n^2 }{2 \sigma^2}} \nonumber \\ & =&
  \frac{1}{\ln(10)\sqrt{2\pi \sigma^2} }\int_{0}^{\infty}dz_n\left (\frac{1}{\mu+z_n/\sqrt {n}}+\frac{1}{\mu-z_n/\sqrt {n}}\right) \exp^{-\frac{z_n^2 }{2 \sigma^2}} \nonumber\\ &= & \frac{1}{\ln(10)\sqrt{2\pi \sigma^2} }\int_{0}^{\infty}dz_n\left (\frac{2 \mu}{\mu^2-z_n^2/n }\right)\exp^{-\frac{z_n^2 }{2 \sigma^2}}.
\end{eqnarray}

Next, considering $n$ sufficiently large such as $(\sigma/\mu)^2 \ll n$, we can perform a series expansion of the term 
$z_n^2/n$ in the denominator of the integrating function, and  the above equation can be approximated as

\begin{equation}
  \langle b_n \rangle=
  \frac{1}{\ln(10)\sqrt{2\pi \sigma^2} }\frac{2}{\mu}
  \int_{0}^{\infty}dz_n\left (1  +\frac{z_n^2}{\mu^2n}\right)\exp^{-\frac{z_n^2 }{2 \sigma^2}}+ {\cal O}\left(1/n^2\right)
\end{equation}

leading to

\begin{equation}
  \langle b_n \rangle=
  \frac{1}{\ln(10)}\left(\frac{1}{\mu}+\frac{\sigma^2}{\mu^3n}\right)+ {\cal O}\left(1/n^2\right)
  \label{avebn}
\end{equation}

Taking into account that $\mu=\langle m\rangle-m_{th}$, from Eq.(\ref{ml}), Eq.(\ref{avebn}) can be written as

\begin{equation}
  \langle b_N \rangle=
  \frac{1}{\ln(10)} \frac{1}{\mu} \left(1+\frac{1}{n} CV^2\right),
  \label{avebn3}
\end{equation}
where $CV=\sigma/\mu$ is the variability coefficient, defined as the ratio between the standard deviation and the mean value of the magnitude distribution.

When the magnitude distribution obeys the GR law, the values of $\mu$ and $\sigma$ can be 
easily evaluated 
\begin{equation}
  \mu=b \ln(10) \int_{m_{th}}^{\infty} 10^{-b (m-m_{th})} (m-m_{th}) dm=\frac{1}{\ln(10) b}
    \label{mu}
\end{equation}

and

  \begin{equation}
\sigma^2=b \ln(10) \int_{m_{th}}^{\infty} 10^{-b (m-m_{th})} (m-m_{th})^2 dm-\mu^2=\left (\frac{1}{\ln(10) b}\right)^2,
  \label{sigma}
\end{equation}
  leading to $CV=1$.

Using the above expressions in Eq.(\ref{avebn}) we finally obtain
\begin{equation}
  \langle b_N \rangle=b \left(1+\frac{1}{n}\right),
    \label{avebn2}
\end{equation}
which, since $b_{\infty}=b$, coincides with Eq.(\ref{oy}) at first order in $(1/n)$.  
We remark that Eq.(\ref{avebn3}) is more general than Eq.(\ref{oy}), or equivalently Eq.(\ref{avebn2}), since it also holds in the case that the magnitude distribution does not follow the GR law. In the next subsection we discuss how one can use Eq.(\ref{avebn}) to quantify deviations from the GR law.   

\subsection{Estimate of the completeness magnitude from \lowercase{$\langle b_n \rangle$}  }\label{sec22}

For an exponential distribution $CV=1$, as evident from Eq.s(\ref{mu},\ref{sigma}), and the estimate of $CV$ is a standard tool to identify deviations from exponential distribution of intertimes and therefore to characterize deviations from a Poisson process in time \cite{BLGdA09}.
Here we propose the use of Eq.(\ref{avebn3}) to identify the completeness magnitude $m_c$.
As anticipated in the introduction, $m_c$ is generally identified by finding the minimum magnitude above which one can neglect deviations from a pure exponential decay of the magnitude distribution. Using this definition of $m_c$, restricting to events with magnitude $m \ge m_{th} \ge m_c$ in the evaluation of both $\mu$ and $\sigma$, one has $CV=1$ and Eq.(\ref{avebn3}) coincides with Eq.(\ref{avebn2}).
Therefore, only when $m_{th} \ge m_c$, the plot of $\langle b_n(m_{th}) \rangle$ versus $1/n$ forms a straight line with an intercept $\alpha$ and a slope $\beta$, both of which are equal to the true $b$ value.
Accordingly, considering the plot of $\langle b_n (m_{th}) \rangle$ versus $1/n$, for increasing values of $m_{th}$, one expects that when $m_{th} \ge m_{c}$ the parametric plot collapse on the asymptotic curve Eq.(\ref{avebn2}). This provides a novel tool which allows us to identify $m_c$. {
  More precisely, we consider increasing values of $m_{th}$ with a constant incremental step $\Delta m_{th}$. At each value of $m_{th}$, we examine the behavior of $\langle b_n(m_{th}) \rangle$ versus $1/n$ and employ the least squares method to determine the inercept $\alpha(m_{th})$ and slope $\beta(m_{th})$. The value of $m_c$ is determined as the smallest value of $m_{th}$ for which $\alpha(m_{th}) = \beta(m_{th}) = \alpha(m_{th}+\Delta m_{th}) = \beta(m_{th}+\Delta m_{th})$ within statistical uncertainty. To ensure the consistency of the method, we verify that the values of $\alpha(m_{th})$ and $\beta(m_{th})$ remain constant for $m_{th} \ge m_c$. Additionally,  to better highlight deviations from the GR law, we introduce the following quantity}
\begin{equation}
  y_n(m_{th})=(\langle b_n(m_{th}) \rangle/b_{\infty}-1)n,
\label{yn}
\end{equation}
where $b_{\infty}=\alpha(m_{th})$ is the value of $\langle b_n(m_{th}) \rangle$ extrapolated in the limit $n\to \infty$.
From Eq.(\ref{avebn3}) we expect that $y_n$  coincides with $CV^2$, i.e.  $y_n(m_{th})=CV^2$, and, accordingly, $y_n(m_{th})=1$ as soon as $m_{th} \ge m_c$. {The visual inspection of deviations of $y_n(m_{th})$ from $1$ can be considered as a residual diagnostic for assessing the quality of our estimate of $m_c$ and identifying potential issues in the model or experimental data.}

\section{Test of analytical results}

\subsection{Numerical test of Eq.(\ref{avebn2})}

To test the validity of Eq.(\ref{avebn2}) we simulate numerical catalogs containing $n_c=10^6$ magnitudes $m_i\ge m_{th}=1.5$  extracted from the GR distribution (Eq.(\ref{gr})) implementing, for each catalog, a different value of $b$ ranging from $0.6$ to $1.4$, with a step of $0.2$. For each catalog we sample $n$ events, repeating the sampling $10^5$ times, and evaluate the average $b$ value $\langle b_n\rangle$ and its distance to the ``true'' value $\delta b_n=\langle b_n\rangle-b$. We find (Fig.\ref{db}(a)) that $\delta b_n$ is a decreasing function of $n$ converging to zero, for sufficiently large $n$, {as already shown in Fig.1 of \cite{NMZAQ16}}.
The error bar in Fig.\ref{db}(a) represents the standard deviation given by $\sqrt{\left(\langle b_n^2(m_{th}) \rangle-\langle b_n(m_{th}) \rangle^2\right)}/\sqrt{n_{ind}}$, where $n_{ind}=n_c/n$ represents the number of independent subsets containing $n$ event in a set with $n_c$ total events. This definition of the error bar, which results to be substantially independent of $n$, is implemented also for the other figures of the manuscript. At the same time we verify that $\delta b_n=\langle b_n \rangle-b_{\infty}$ is in an excellent agreement with the correcting term $\delta b_n=b_{\infty}(1/n)$ as predicted by Eq.(\ref{avebn2}), for all $b$ values (Fig.\ref{db}(a)).

\subsection{Testing for instrumental catalogs}

In this article we analyze three different earthquake catalogs for three different geographical regions: Southern California, Japan and New Zealand.
The Southern California seismic catalog, fixing $m_{th}=2.5$, contains 47713 events in a period spanning from 1981/01/01 to 2019/12/31. The Japanese JMA catalog contains 3038158 events considering the time period 2000/01/01-2018/12/31, with $m_{th}=1.7$
. Finally the seismic catalog of New Zealand contains 284612 events with $m_{th}=1.7$, in a period spanning from 1460/01/01 to 2020/12/31. \\
The Websites where the catalogues can be downloaded are reported in the Data availability section.

\begin{figure}
\includegraphics[width=17.1cm]{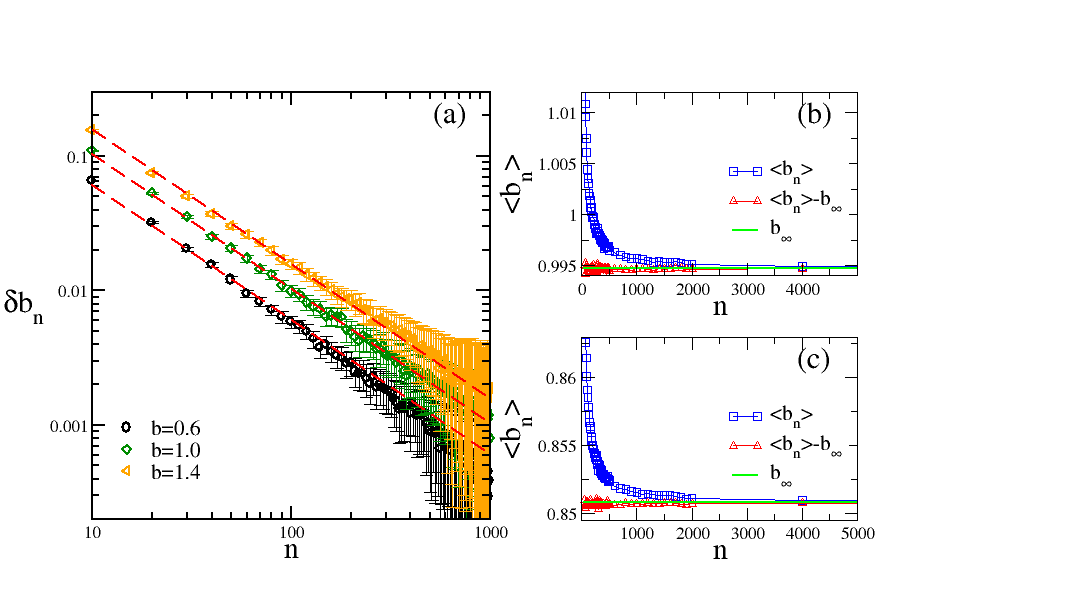}
\caption{(Panel a) The distance $\delta b_n=\langle b_n \rangle-b_{\infty}$ from the asymptotic $b_{\infty}$ value as a function of $n$, for different choices of $b_{\infty}=b$ in simulated catalogs. The red dashed lines represent the theoretical expression $\langle b_n\rangle=b(1-n^{-1})$. (Panel b) The value of $\langle b_n \rangle$ for the Southern California catalog (black circles) is compared with $\langle b_n \rangle-b_{\infty}(1/n)$ (red diamonds) and with $b_{\infty}$ (green continuous line). Here the completeness magnitude has been fixed at $m_{th}=2.5$. 
(Panel c) The equivalent of panel (b) for the Japanese (JMA) considering $ 3038158 $ events with magnitude $m_{th}>1.7$.}
\label{db}
\end{figure}

In Fig.\ref{db}(b) we test the validity of  Eq.(\ref{avebn2})
in instrumental catalogs, by considering the relocated Southern California and the Japanese JMA catalog. Similar results, not plotted, are obtained for the New Zealand catalog.  
More precisely, for each value of $n$ (with $n$ varying from $50$ to $500$ with a step $10$ and from $500$ to $2000$ with a step $100$, then $n$ assumes the values $4000$, $ 7000$ and $10000$) we have considered $10^5$ randomly choosing subsets of the catalog, each composed of $n$ earthquakes, and we have evaluated $\langle b_n \rangle$.
Results plotted in Fig.\ref{db}b and Fig.\ref{db}c show that {deviations of $\langle b_n \rangle-b_{\infty}(1/n)$ from the true asymptotic value is less than $10^{-4}$ for $n=50$, and indeed, the two curves are indistinguishable for all the $n$ values considered. Conversely, without incorporating the correcting term, to achieve a similar accuracy, we must consider a dataset including at least $n\gtrsim 1400$ earthquakes.}  
This result indicates that taking properly into account the correcting term one can achieve a more reasonable estimate of the asymptotic $b$ value, when $n$ is finite.

\subsection{Test of the method for measuring the completeness magnitude from \lowercase{$\langle b_n \rangle$} }

\begin{figure}
  \includegraphics[width=18cm]{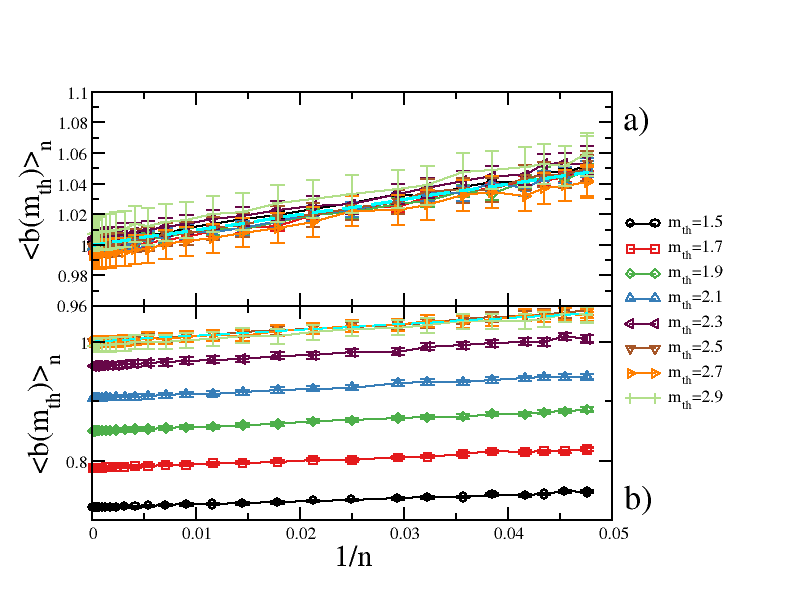} 
\caption{(a) $\langle b_n(m_{th}) \rangle$ is plotted versus $1/n$ for a complete numerical catalog.
  Different curves correspond to different values of $m_{th}$ (see legend). The { cyan} dashed line is the theoretical prediction $\langle b_n(m_{th}) \rangle =b(1+1/n)$.
  (b) The same study of panel (a) for an incomplete numerical catalog, with completeness magnitude $m_{th}=2.5$.}
\label{figbm1}
\end{figure}

\begin{figure*}
  \includegraphics[width=14.cm]{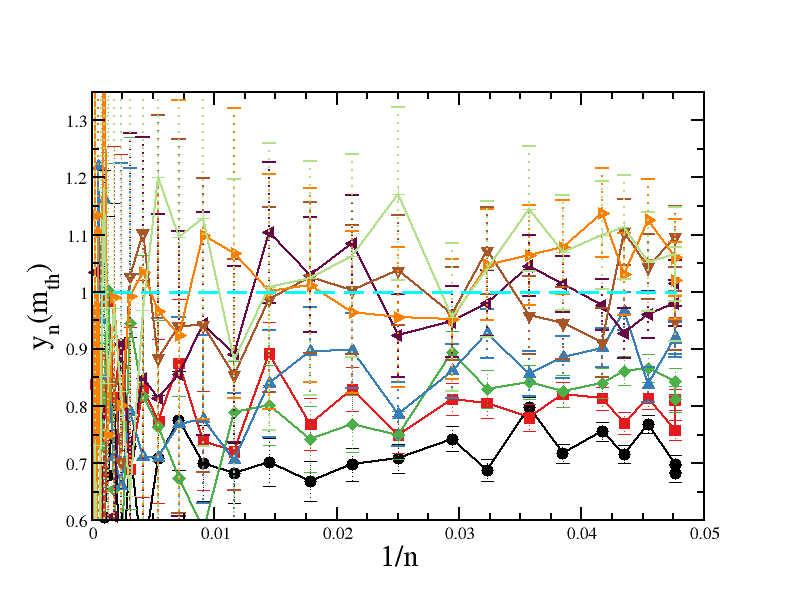}
\caption{The quantity  $y_n(m_{th})$ versus $1/n$ for the same data sets considered in panel (b) of Fig. \ref{figbm1}, with the same color code. The {cyan} dashed line is the asymptotic prediction  $y_n(m_{th})=CV^2=1$, for $m \ge m_{th}$. } 
\label{figbm1c}
\end{figure*}

In Fig.\ref{figbm1} and Fig.\ref{figbm1c} we present the tool described in Sec.\ref{sec22} applied to numerical catalogs. More precisely we start by considering a complete catalog containing $10^6$ earthquakes with magnitude $m\ge 1.5$ distributed according to the GR law with $b=1$. We set $\Delta m_{th}=0.2$ and consider different magnitude thresholds $m_{th}=1.5+k\Delta m_{th}$ with $k \in [0,8]$ and for each value of $m_{th}$ we consider $10^5$ randomly chosen subsets of the catalog, each composed of $n$ earthquakes, and we evaluate $\langle b_n(m_{th}) \rangle$. The plot of $\langle b_n(m_{th}) \rangle$ versus $1/n$ is presented in Fig.\ref{figbm1}a which shows that, within numerical uncertainty, curves with the different $m_{th}$ collapse on the same master curve which, as expected, is consistent with Eq.(\ref{avebn2}), $\langle b_n(m_{th}) \rangle=b(1+1/n)$, with $b=1$. This is confirmed by the least squares method which correctly gives $m_c=1.5$.

In Fig.\ref{figbm1}b we present the same study applied to the same numerical catalog but taking into account incompleteness by assuming that not all small earthquakes are reported in the catalog. For this reason we remove an earthquake with magnitude $m$ from the original complete catalog with a probability $Q(m)$ which is larger the smaller  is $m$. In particular we use  $Q(m)=(2/3)(2.5-m)$ for $m \le 2.5$, which implies that the $66 \%$ of earthquakes with magnitude $m=1.5$ are, on average, removed from the catalog, whereas we assume $Q(m)=0$ for $m\ge 2.5$, which implies that no earthquake with magnitude $m \ge 2.5$ is removed from the catalog. Accordingly the completeness magnitude of our numerical catalog is $m_c=2.5$.  
The plot of $\langle b_N(m_{th}) \rangle$ versus $1/n$ (Fig.\ref{figbm1}b) shows that curves clearly depend on $m_{th}$ for $m_{th}< 2.5$, whereas they collapse to $\langle b_n(m_{th}) \rangle=1+1/n$ for $m_{th} \ge 2.5$. {Indeed, curves with $m_{th} \ge 2.5$ are indistinguishable in Fig. \ref{figbm1}b.} This is confirmed by the least squares method which in this case gives $m_c=2.5$.
In order to enlighten deviations from the GR law in Fig.\ref{figbm1c} we plot $y_n(m_{th})$, defined in Eq.(\ref{yn}), versus $1/n$ and we find that $y_n(m_{th})$, for any $n$ value,  monotonically increases with $m_{th}$ and saturates to $1$ for $m_{th} \ge 2.5$. Accordingly, this analysis provides a $CV$ value smaller than $1$ when $m_{th}<2.5$, correctly identifying deviations from the GR law and, at the same time, providing the correct value of the completeness magnitude $m_c=2.5$.                

\begin{figure}
\vskip -3.5cm
  \includegraphics[width=20cm]{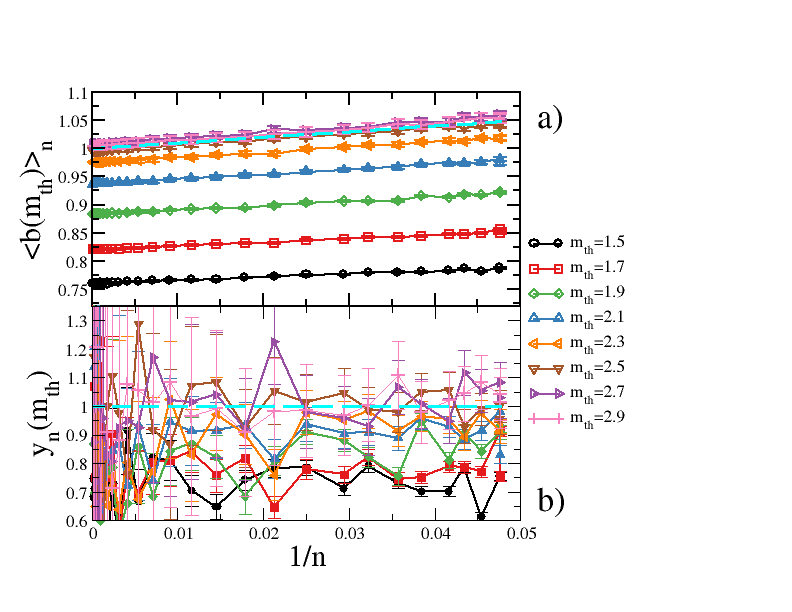}
  \caption{(a) $\langle b_n(m_{th}) \rangle$ is plotted versus $1/n$ for the Southern California catalog.
    Different curves correspond to different values of $m_{th}$ (see legend). The dashed cyan line is the theoretical prediction $\langle b_n(m_{th}) \rangle =b(1+1/n)$. (b) The quantity  $y_n(m_{th})$ versus $1/n$ for the same data sets considered in panel (a). The dashed cyan line is the asymptotic prediction  $y_n(m_{th})=1$.} 
\label{figbm2}
\end{figure}

\begin{figure}
  \vskip -3.5cm
  \includegraphics[width=20cm]{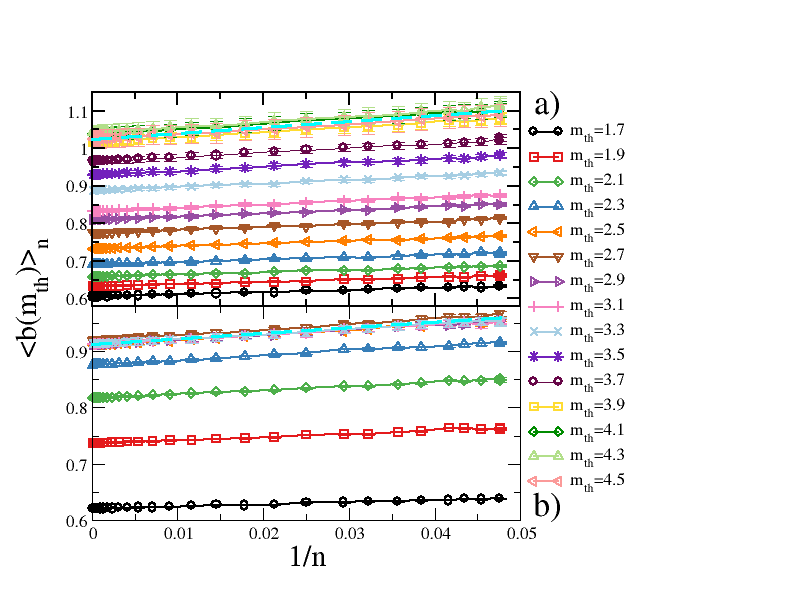}
  \caption{$\langle b_n(m_{th}) \rangle$ is plotted versus $1/n$ for the Japan Catalog (a) and for the New Zealand catalog (b). Different curves correspond to different values of $m_{th}$ (see legend). The dashed cyan line, in each panel, is the theoretical prediction $\langle b_n(m_{th}) \rangle =b(1+1/n)$.}
\label{figbm3}
\end{figure}

\begin{figure}
  \vskip -3.5cm
    \includegraphics[width=20cm]{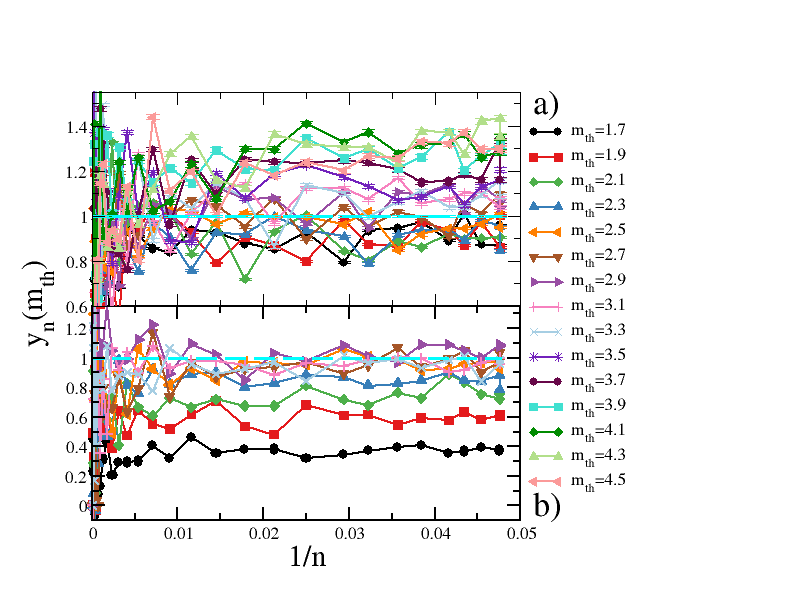}
  \caption{ $y_n(m_{th})$ is plotted versus $1/n$ for the Japan Catalog (a) and for the New Zealand catalog (b). Different curves correspond to different values of $m_{th}$ (see legend). The dashed cyan line is the theoretical prediction  $y_n(m_{th})=CV^2=1$.}
\label{figbm4}
\end{figure}

  In Fig.\ref{figbm2} and Fig.\ref{figbm3} we present the analogous analysis of Fig.\ref{figbm1} for the instrumental catalogs of Southern California, Japan and New Zealand. We find that the plot $\langle b_n(m_{th}) \rangle$ versus $1/n$ is very similar to the one observed in the uncomplete simulated catalog (Fig.\ref{figbm1}b) for all the three catalogs. In particular we observe that $\langle b_n(m_{th}) \rangle$, for different $m_{th}$ in Southern California collapse on the same master curve for $m_{th}>2.5$. Our method in this case gives $m_c=2.5$ and $b_{\infty}=0.994\pm 0.02$. This estimate of $m_c$ is supported by Fig.\ref{figbm2}b where we present the plot of $y_n(m_{th})$ versus $1/n$, for the instrumental catalogs of Southern California. We find that for $m_{th}>2.5$ $y_n \simeq 1$ indicating that the magnitude distribution is consistent with the GR law and clearly supporting $m_{c}=2.5$ fo Southern California, in agreement with results obtained with other methods (see among the others \cite{GLD14}). In Fig.s \ref{figbm3} and \ref{figbm4} we present the same results for the Japan  and the New Zealand  catalogs. 
  The plot of $\langle b_n(m_{th}) \rangle$ vs $n$ for New Zealand (Fig.\ref{figbm3}b) show a collapse of curves for different $m_{th}$ when $m_{th}=2.5$ whereas the collapse in Japan is recovered at a larger value $m_{th}=3.9$ (Fig.\ref{figbm3}a). Nevertheless, a very important difference between Japan and New Zealand is observed in Fig.\ref{figbm4} where we plot $y_n$ vs $1/n$ for the two catalogs. Indeed the New Zealand catalog presents a pattern very similar to the incomplete numerical catalog (Fig.\ref{figbm2}b) with $y_n \simeq 1$ when $m_{th}>2.5$, clearly indicating $m_c=2.5$ with $b_{\infty}=0.92\pm 0.02$ obtained from Fig.\ref{figbm3}b. Conversely, in the Japan catalog $y_n$ never saturates to $1$ also for the largest value of $m_{th}$ considered. This indicates that, even if $\langle b_n(m_{th}) \rangle$ appears independent of $m_{th}$ when $m_{th}>3.9$, the $CV$ remains significantly different than $1$ indicating important deviations from the GR law. For larger values of $m_{th}>4.5$ the Japan catalog contains less than $10^3$ events and fluctuations in $\langle b_n(m_{th}) \rangle$ becomes to relevant to prevent us to reach any conclusion. We can, therefore, only established that the completeness magnitude for Japan is larger than $m_{th}=4.5$, without providing a precise value, consistently with other analysis performed by \cite{zhuang}.

\section{Conclusions}

We have shown that when a sample of earthquakes is extracted from a simulated catalog, the evaluated $b$ value is 
overestimated by a term inversely proportional to the number of events $n$ used for its evaluation, confirming the result obtained by \cite{OY86}.
We have shown that this systematic effect can be also explained by means of the central limit theorem. 
We have used this result to provide a tool to identify the completeness magnitude $m_c$. The 
method can be considered an extension of the one proposed by \cite{CG02} which defines $m_c$ as the threshold magnitude 
$m_{th}$ above which the estimate of $b$ from Eq.(\ref{ml}) becomes independent of $m_{th}$. According to 
Eq. (\ref{avebn3}), in our approach, the saturation of $b$ is a condition necessary but not sufficient for an exponential 
magnitude distribution. Indeed, the saturation of $b$ only correspond to the request that the mean value of the magnitude 
$\mu$ becomes $m_{th}$ independent. In our approach, conversely, we ask that also the standard deviation must saturate to 
a specific value, leading to $CV=1$. In other words, the method by \cite{CG02} is a condition only on the first momentum 
of the magnitude distribution, whereas our method imposes a constraint also on the second momentum. It can be considered 
alternative to other methods where the constraint is imposed on the full function form of the magnitude distribution as 
for instance in \cite{MSST20, HM21} when the Lilliefors test is applied to assess if the magnitude distribution is 
exponential. {The visual examination of the behavior of $y_n(m_{th})$ at various $m_{th}$ values enables us to evaluate the accuracy of our estimate of $m_c$ and identify potential issues with the experimental data, as demonstrated by the analysis of the Japan seismic catalog.}


\section{Declarations}
The authors declare that there is no conflict of interest.

\section{Ethical statements}
All materials and results here presented have not been previously published elsewhere. The paper properly credits the meaningful contributions of co-authors.

\section{Data availability}
Data here used can be found and downloaded at the web-site \url{https://scedc.caltech.edu/data/alt-2011-dd-hauksson-yang-shearer.html}, \url{https://www.data.jma.go.jp/svd/eqev/data/bulletin/index_e.html} and \url{https://www.geonet.org.nz/data/types/eq_catalogue} for Southern California, Japan and New Zealand, respectively.

\section{acknowledgments}
We would like to acknowledge the Southern California Earthquake Data Center (SCEDC) for proving the earthquake catalog which can be downloaded at the following web site: 
\url{https://scedc.caltech.edu/data/alt-2011-dd-hauksson-yang-shearer.html}, the Japan Meteorological Agency (JMA) for providing the Japanese earthquake catalog at the web site: \url{Japan Meteorological Agency (2018). The seismological bulletin of Japan. https://www.data.jma.go.jp/svd/eqev/data/bulletin/index_e.html} and GNS for providing the New Zealand earthquake catalog at: \url{https://www.gns.cri.nz/data-and-resources/national-earthquake-information-database/}.

C.G. would like to thanks the MIUR project PRIN 2017 WZFT2p for financial support. E.L. acknowledges support from the MIUR PRIN 2017 project 201798CZLJ. G.P. would like to thanks the MEXT Project for Seismology TowArd Research innovation with Data of Earthquake (STAR-E Project), Grant No.: JPJ010217. We also thank Prof.Jiancang Zhuang for helpful information provided.


\bibliographystyle{abbrvnat}


\end{document}